\documentclass[10pt,twocolumn,twoside] {IEEEtran}

\usepackage{psfrag,epsfig,graphics}
\usepackage{amsmath,amssymb,amsthm,multirow}
\usepackage{cite,subfigure,balance}

\usepackage{color}

\def\cwt{\textcolor{white}}
\def\cmag{\textcolor{black}}  
\definecolor{violet}{rgb}{0.5,0,0.5}

\newcommand{\cb}[1]{{\boldsymbol{#1}}}
\newcommand{\cp}[1]{\ifmmode {\mathcal{#1}}\else ${\mathcal{#1}}$\fi}

\newcommand{\bK}{\boldsymbol{K}}

\newcommand{\bQ}{\boldsymbol{Q}}
\newcommand{\bR}{\boldsymbol{R}}

\newcommand{\bX}{\boldsymbol{X}}

\newcommand{\bw}{\boldsymbol{w}}
\newcommand{\btw}{\tilde{\boldsymbol{w}}}

\newcommand{\bx}{\boldsymbol{x}}

\newcommand{\tr}{\text{tr}}

\newcommand{\bSig}{\boldsymbol{\Sigma}}

\newcommand{\E}{{\mathbb{E}}}

\newcommand{\sgn}{\text{sgn}}

\usepackage{tikz}

\newtheorem{lemma}{Lemma}

\begin{document}

\title{Transient Performance Analysis \\ of Zero-Attracting LMS}

\author{Jie Chen, \emph{Member, IEEE},\; C{\'e}dric Richard, \emph{Senior Member, IEEE}, \; Yingying Song, \;David Brie, \emph{Member, IEEE}

\thanks{The work of J. Chen was supported in part by NSFC grants 61671382 and 61671380. The work of C. Richard was supported in part by ANR and DGA grant ANR-13-ASTR-0030 (ODISSEE project). The work of Y. Song and D.~Brie were supported by the FUI AAP 2015 Trispirabois Project, the Conseil R\'egional de Lorraine and the GDR ISIS.}
\thanks{J. Chen is with Centre of Intelligent Acoustics and Immersive Communications at School of Marine Science and Technology, Northwestern Polytechinical University, Xi'an, China (email: dr.jie.chen@ieee.org). C. Richard is with the Universit\'e C\^ote d'Azur, OCA, CNRS, France (email: cedric.richard@unice.fr), and on leave at INRIA Sophia Antipolis, France. Y.~Song and D.~Brie are with Centre de Recherche en Automatique de Nancy (CRAN), Universit\'e de Lorraine, CNRS (email: firstname.lastname@univ-lorraine.fr).}}

\maketitle

\begin{abstract}
Zero-attracting least-mean-square (ZA-LMS)  algorithm has been widely used for online sparse system identification. It combines the LMS framework and $\ell_1$-norm regularization to promote sparsity, and relies on subgradient iterations. Despite the significant interest in ZA-LMS, few works analyzed its transient behavior. The main difficulty lies in the nonlinearity of the update rule. In this work, a detailed analysis in the mean and mean-square sense is carried out in order to examine the behavior of the algorithm. Simulation results illustrate the accuracy of the model and highlight its performance through comparisons with an existing model.
\end{abstract}


\section{Introduction}

The need for finding sparse solutions to system identification problems has been motivated by many applications in signal and image processing~\cite{bruckstein2009sparse,Elad2010sparse}, such as those in biomedical imaging~\cite{fessler2010model}, remote sensing~\cite{iordache2011sparse} and communications~\cite{cotter2002sparse}. For instance, channel estimation problems have emphasized the need for sparsity-aware algorithms because channels with a sparse impulse response arise in a number of practical situations. In all cases, the underlying problem can be expressed in terms of finding the $\ell_0$-based sparsest solution, i.e., to find a minimal number of non-zero coefficients that represent the solution of interest. Despite the problem being generally NP-hard, compressive sensing  theory has been developed to provides a robust framework to find sparse solutions with low computational complexity. Regularization with $\ell_1$-norm is popularly used as a surrogate of the non-convex $\ell_0$-norm to promote the sparsity~\cite{candes2008introduction}.

Built on this principe, zero-attracting LMS (ZA-LMS) was proposed in~\cite{Chen2009ZALMS}. The algorithm combines the LMS framework and $\ell_1$-norm regularization, and relies on subgradient iterations. Its stability was analyzed in~\cite{Chen2009ZALMS,Zhang2014}. Despite the significant interest in ZA-LMS, few works analyzed its behavior. The main difficulty lies in the nonlinearity of the update rule, which arises from the subgradient of the $\ell_1$-norm term. Most related works include~\cite{Zhang2014} and~\cite{Shi2010}. In~\cite{Shi2010}, the analysis of the transient behavior of ZA-LMS is limited to white input signals. The proposed model involves terms that cannot be explicitly evaluated. In~\cite{Zhang2014}, the authors propose a more comprehensive analysis. This work evaluates the mean weight behavior in an exact manner under the Gaussian assumption on estimated weights (or, equivalently, weight errors). Nevertheless, coarse approximations are adopted for the mean-square error analysis. They consist of simply substituting the expectation of the product of two sign functions by the product of their expectations. Finally, another work~\cite{Su2012} suggests an analysis of the $\ell_0$ regularized LMS by classifying the weights into distinct classes according to their values.

The main difficulty in the analysis of ZA-LMS behavior lies in evaluating expected values that involve the sign of non-zero mean random variables. Making assumptions which are consistent with~\cite{Chen2009ZALMS,Shi2010,Zhang2014,Su2012,Chen2014NNLMS}, we derive an exact model of the transient behavior of ZA-LMS without further approximations. Simulations illustrate the consistency between the simulated results and the theoretical findings, as well as the improved accuracy compared with the previous works. 

\smallskip

\noindent \textbf{Notation. } Boldface small letters $\bx$ denote column vectors. Boldface capital letters $\cb{X}$ denote matrices. $[\bx]_i$ and $[\bX]_{ij}$ denote respectively the $i$-th entry of $\bx$, and the $(i,j)$-th entry of $\cb{X}$. The superscript $(\cdot)^\top$ denotes the transpose of a matrix or a vector.  All-zero vector of length $N$ is denoted by $\cb{0}_N$. The operator $\sgn\{\cdot\}$ takes the sign of the entries of its argument.  The operator $\text{tr}\{\cdot\}$ takes the trace of its matrix argument. 
The Gaussian distribution with mean $\mu$ and variance $\sigma^2$ is denoted by  $\cp{N}(\mu,\sigma^2)$. In the multivariate case, it is denoted by $\cp{N}(\cb{\mu},\bSig)$. The cumulative distribution function (CDF) of the standard Gaussian distribution is denoted by $\phi(x)$. The CDF of the multivariate Gaussian distribution with mean $\cb{\mu}$ and covariance $\bSig$ is denoted by $\Phi(\bx, \cb{\mu}, \bSig)$.

\section{Problem formulation and  ZA-LMS }

Consider an {unknown system} with input-output relation characterized by the linear model:
\begin{equation}
	\label{eq:linear.model}
	y_n = \bx_n^\top\bw^{\star} + z_n
\end{equation}
with $\bw^\star \in \mathbb{R}^L$ an unknown parameter vector, and $\bx_n \in \mathbb{R}^L$ the regressor \cmag{with correlation matrix $\bR_x > 0$} at instant $n$. The regressor $\bx_n$ and the output signal $y_n$ are assumed to be zero-mean and stationary. The modeling error $z_n$ {is assumed to be stationary, independent and identically distributed (i.i.d.), with zero-mean and variance $\sigma_z^2$, and independent of any other signal}. We consider the system identification problem:
\begin{equation}
         \label{eq:mse.nng}
         \begin{split}
                    \bw^o =& \arg\min_{\bw} \, \mathbb{E}\big\{[y_n - \bw^\top\bx_n]^2 \big\} + \lambda \|\bw\|_1  \\
         \end{split}
\end{equation}
where the $\ell_1$-norm is particularly useful for applications where the parameter vector has a sparse structure. In order to solve this regularized problem, zero-attracting LMS was proposed in~\cite{Chen2009ZALMS} based on the subgradient iteration:
\begin{equation}
           \label{eq:update}
            \bw_{n+1} = \bw_{n} + \mu \, e_n  \bx_n - \rho \, \sgn\{\bw_n\}
\end{equation}
where $e_n=y_n-\bw_n^\top\bx_n$, $\mu$ is a positive step size, and  $\rho =\mu\lambda$.

\section{Transient behavior model of ZA-LMS}
%

We now study the transient behavior of the ZA-LMS algorithm. Defining the weight error vector $\btw_n$ as the difference between the estimated weight vector $\bw_n$ and $\bw^\star$, namely,
\begin{equation}
        \label{eq:tw.def}
        \btw_n  = \bw_n - \bw^\star,
\end{equation}
the analysis of ZA-LMS consists of studying the evolution of  the first and second-order moments of $\btw_n$ over time.

\subsection{Statistical assumptions}
We introduce the following statistical assumptions to keep the calculations mathematically tractable:
\begin{itemize}
	\item [\textbf{A1:}] The weight-error vector $\btw_n$ is statistically independent of the input vector $\bx_n$. 
	\item [\textbf{A2:}] Any pair of entries $[\btw_n]_i$ and $[\btw_n]_j$ with $i\neq j$ is jointly Gaussian. 
\end{itemize}
The independence assumption A1 is widely used in the analysis of adaptive filters~\cite{haykin2005,sayed2008adaptive}. Assumption A2 is consistent with the Gaussian assumptions in~\cite{Chen2009ZALMS,Shi2010,Su2012,Chen2014NNLMS,Zhang2014}. It is weaker than in~\cite{Chen2009ZALMS,Shi2010,Su2012,Chen2014NNLMS} which assume that $\btw_n$ is Gaussian distributed, but stronger than in~\cite{Zhang2014} which assumes that the entries $[\btw_n]_i$ are Gaussian random variables. Nevertheless, the authors in~\cite{Zhang2014} use approximations that are ideally suited for independent entries $[\btw_n]_i$. We illustrate assumption A2 in~Fig.~\ref{fig:hist} with some histograms of $[\btw_n]_i$ versus $[\btw_n]_j$. We shall show it is sufficient to make the calculation of the nonlinear terms tractable without further approximations, and to derive a more accurate model.

\subsection{Mean weight behavior model}
We first focus on the mean weight behavior of the algorithm. Subtracting $\bw^\star$ from both sides of~\eqref{eq:update}, and using $e_n = z_n - \btw_n^\top \bx_n$, yields the update relation of $\btw_n$: 
\begin{equation} 
       \label{eq:btw.update}
       \btw_{n+1} = \btw_n - \mu \, \bx_n\bx_n^\top \btw_n + \mu \, z_n \bx_n - \rho\, \sgn\{\bw^\star+\btw_n\}
\end{equation}
Taking the expectation of~\eqref{eq:btw.update} and considering A1,  we have:
\begin{equation} 
       \label{eq:Ebtw.update1}
       \E\{ \btw_{n+1} \} = \E\{\btw_n\} \!-\! \mu \, \bR_x \E\{\btw_n\}  \!-\! \rho\, \E\{ \sgn\{\bw^\star+\btw_n\}\}.
\end{equation}
To characterize the evolution of $\E\{ \btw_{n+1} \}$ it is necessary to evaluate the last term  $\E\{ \sgn\{\bw^\star+\btw_n\}\}$ in~\eqref{eq:Ebtw.update1}.

\noindent{\lemma Consider a random variable $u\sim\cp{N}(\mu, \sigma^2)$. The expectation of its sign value is given by:
\begin{equation}
	\label{eq:lemma1}
	\E\{\sgn\{u\}\} = 1 - 2  \phi(-{\mu}/{\sigma})
\end{equation}
\qed}
\smallskip

\noindent As shown in Appendix~A, proving this result is almost trivial. Then, the entries of $\E\{ \sgn\{\bw^\star+\btw_n\}\}$ are obtained by making the following identification:
 \begin{equation}
         u \leftarrow [\bw^\star+\btw_n]_i
 \end{equation}
 with 
 \begin{align}
         \mu & \leftarrow  [\bw^\star]_i+\E\{[\btw_n]_i\}   \\
         \sigma^2 &\leftarrow \E\{[\btw_n]_i^2\} - \E\{[\btw_n]_i\}^2
 \end{align}
 where $\E\{[\btw_n]_i^2\}$ can be extracted from the diagonal of the second-order moment matrix $\bK_n = \E\{\btw_n\btw_n^\top\}$ that will be determined in the next subsection.

 \subsection{Mean-square error behavior model}

Using $e_n = z_n - \btw_n^\top\bx_n$ and considering A1 leads to the following expression for the mean-square error (MSE)~\cite{sayed2008adaptive}:
\begin{equation}
	\label{eq:MSE}
	\E\{e^2_n\} \approx \sigma_z^2  +\tr\{\bR_x\bK_n\}
\end{equation}
where $\tr\{\bR_x\bK_n\}$ is the excess mean-square error (EMSE). We thus need to determine a recursion for $\bK_n$ in order to evaluate the MSE (or EMSE). Post-multiplying~\eqref{eq:btw.update} by its transpose, taking the expected value, and using A1 leads to: 
 \begin{equation}
	\label{eq:K.update}
	\begin{split}
	\bK_{n+1} 
		&=\bK_n+ \mu^2\sigma_z^2\bR_x + \mu^2\bQ_1 + \rho^2 \bQ_2\\
		&-\mu (\bQ_3+\bQ_3^\top) - \rho (\bQ_4+\bQ_4^\top) +\mu\rho (\bQ_5+\bQ_5^\top)
	\end{split}
\end{equation}
where:
 \begin{align}
	\bQ_1 &= \E\{\bx_n\bx_n^\top\btw_n\btw_n^\top\bx_n\bx_n^\top\}		\label{eq:Q1}\\
	\bQ_2 &= \E\{\sgn\{\bw^\star+\btw_n\}\sgn^\top\!\{\bw^\star+\btw_n\}\}	\label{eq:Q2}\\
	\bQ_3 & = \E\{\btw_n\btw_n^\top \bx_n\bx_n^\top\}					\label{eq:Q3}\\ 
	\bQ_4 & = \E\{\btw_n\sgn^\top\!\{\bw^\star+\btw\}\}					\label{eq:Q4}\\
	\bQ_5 & = \E\{\bx_n\bx_n^\top\btw_n\sgn^\top\!\{\bw^\star+\btw_n\}\}.	\label{eq:Q5}
 \end{align}
We shall now calculate the terms $\bQ_i$ defined in~\eqref{eq:Q1}--\eqref{eq:Q5}. Assumption A1 allows to separate $\bx_n$ and $\btw_n$ in~\eqref{eq:Q1} to help in evaluating $\bQ_1$. No closed-form expression can, however, be provided for if no distribution for input data $\bx_n$ is assumed. In order to evaluate the fourth-order moments of $\bx_n$, and then provide a closed-form expression for $\bQ_1$, we consider the special case where $\bx_n$ is zero-mean Gaussian. Using A1 and Isserlis' theorem (decomposition of higher-order moments of the multivariate Gaussian distribution), we then obtain:
\begin{equation}
	\bQ_1 = 2\bR_x \bK_n \bR_x + \tr\{\bR_x\bK_n\}\bR_x
\end{equation}
The rest of the analysis, which consists of calculating the terms $\bQ_2$ to $\bQ_5$ in~\eqref{eq:Q2}--\eqref{eq:Q5}, remains valid even if the Gaussian assumption on $\bx_n$ is relaxed. To evaluate $\bQ_2$, we introduce the following lemma.
 
\noindent {\lemma Consider two random variables $u$ and $v$ which are jointly Gaussian, namely,
\begin{equation}
	\begin{bmatrix}u \\ v\end{bmatrix} \sim \cp{N} \left( \cb{\mu}:=\begin{bmatrix}\mu_u \\ \mu_v \end{bmatrix} , \, {\bSig}_{uv}:= {\footnotesize  		\begin{bmatrix} \sigma_u^2 & \rho_{uv} \\ \rho_{uv} & \sigma_v^2 \end{bmatrix}} \right) 
\end{equation}
with $\cb{\mu}$ and ${\bSig}_{uv}$ their mean vector and covariance matrix, defined as described above.
The expectation $\E\{\sgn\{u\}\sgn\{v\}\}$ is given by:
\begin{equation}
	\begin{split}
	&\E\{\sgn\{u\}\sgn\{v\}\} \\
	&=
	\Phi(\cb{0}_2, [\mu_u, \mu_v]^\top, \bSig_{uv})+ \Phi(\cb{0}_2, -[\mu_u, \mu_v]^\top, \bSig_{uv}) \\
	&-  \Phi(\cb{0}_2, [\mu_u, -\mu_v]^\top, \overline{\bSig}_{uv}) -  \Phi(\cb{0}_2, [-\mu_u, \mu_v]^\top, \overline{\bSig}_{uv})
       \end{split}
\end{equation}
with $\overline{\bSig}_{uv} = {\bSig}_{uv} \circ {\footnotesize \begin{bmatrix}  \phantom{-}1  & -1\\ -1 & \phantom{-}1 \end{bmatrix}} $, and $\circ$ the element-wise product. \qed}
 \smallskip
 
 \begin{figure*}[!t]
      	\centering
	\subfigure[Weight model ($\lambda = 0.01$)]{\label{fig:weight}
	        
		\includegraphics[trim = 20mm 16mm 25mm 22mm, clip, scale=0.33]{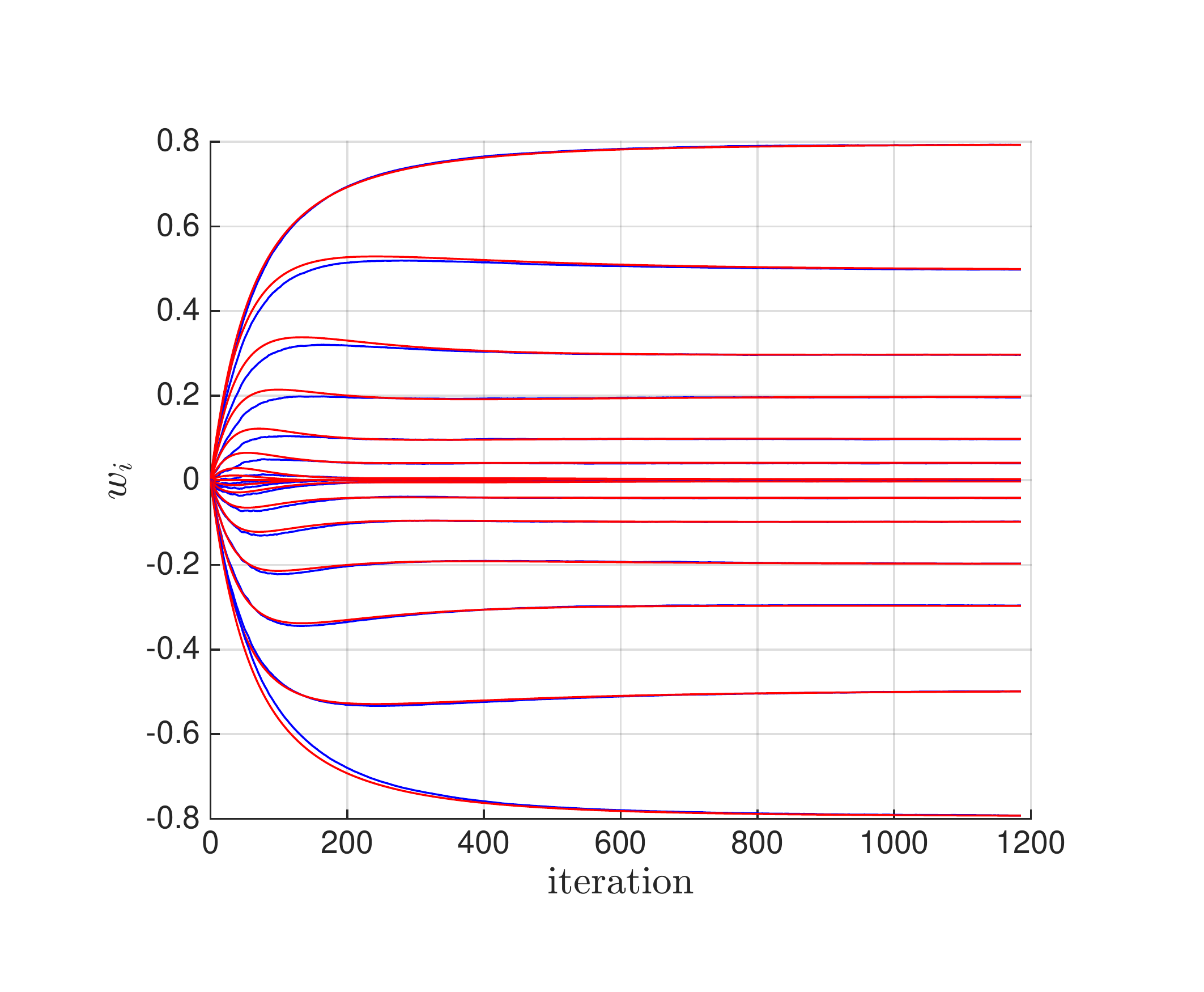} 
          }
          	\subfigure[MSE and EMSE models ($\lambda = 0.01$)]{ \label{fig:mse1}
		\includegraphics[trim = 20mm 16mm 25mm 22mm, clip, scale=0.33]{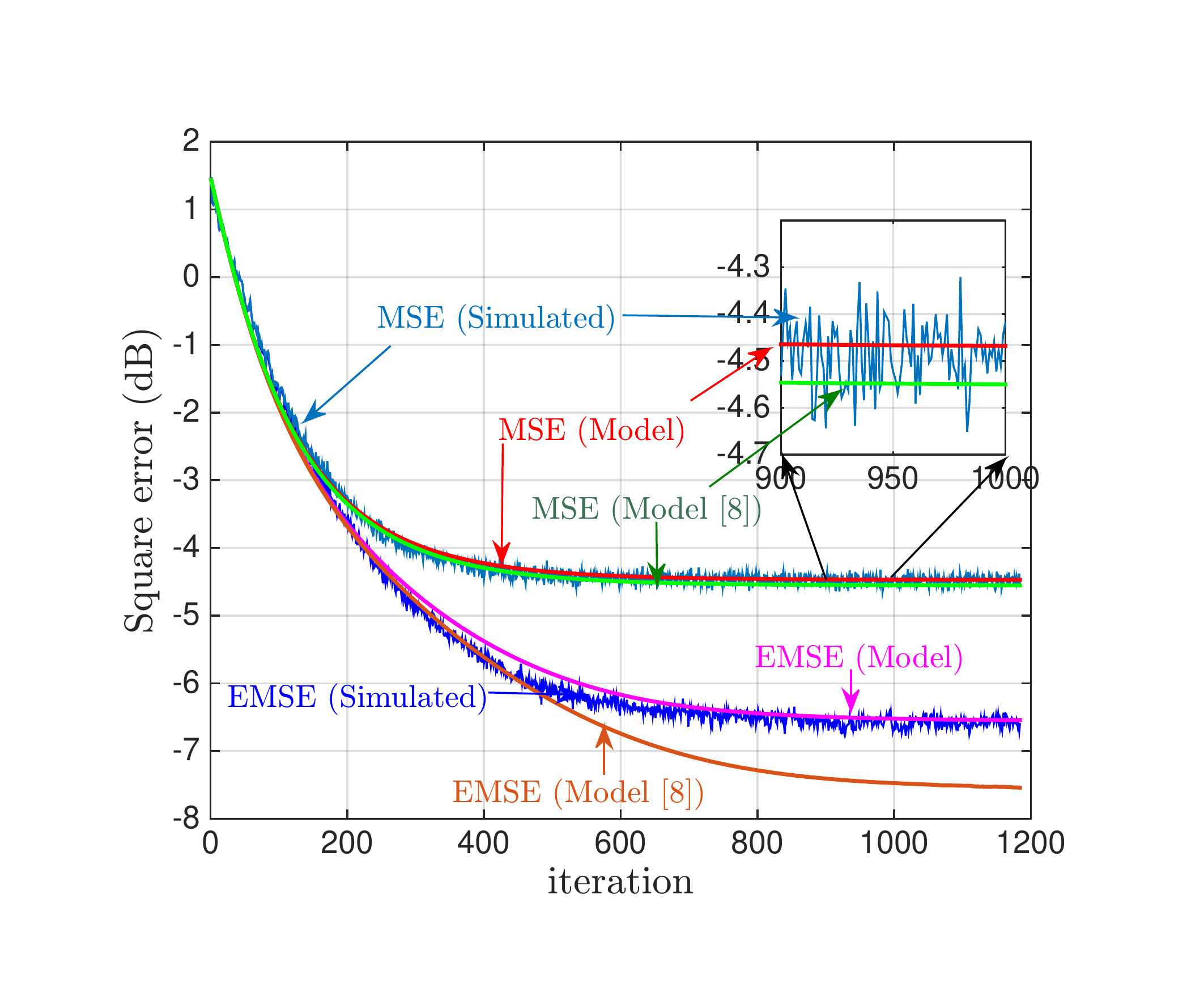} 
          }
          	\subfigure[MSE and EMSE models ($\lambda = 0.001$)]{  \label{fig:mse2}
		\includegraphics[trim = 20mm 16mm 25mm 22mm, clip, scale=0.33]{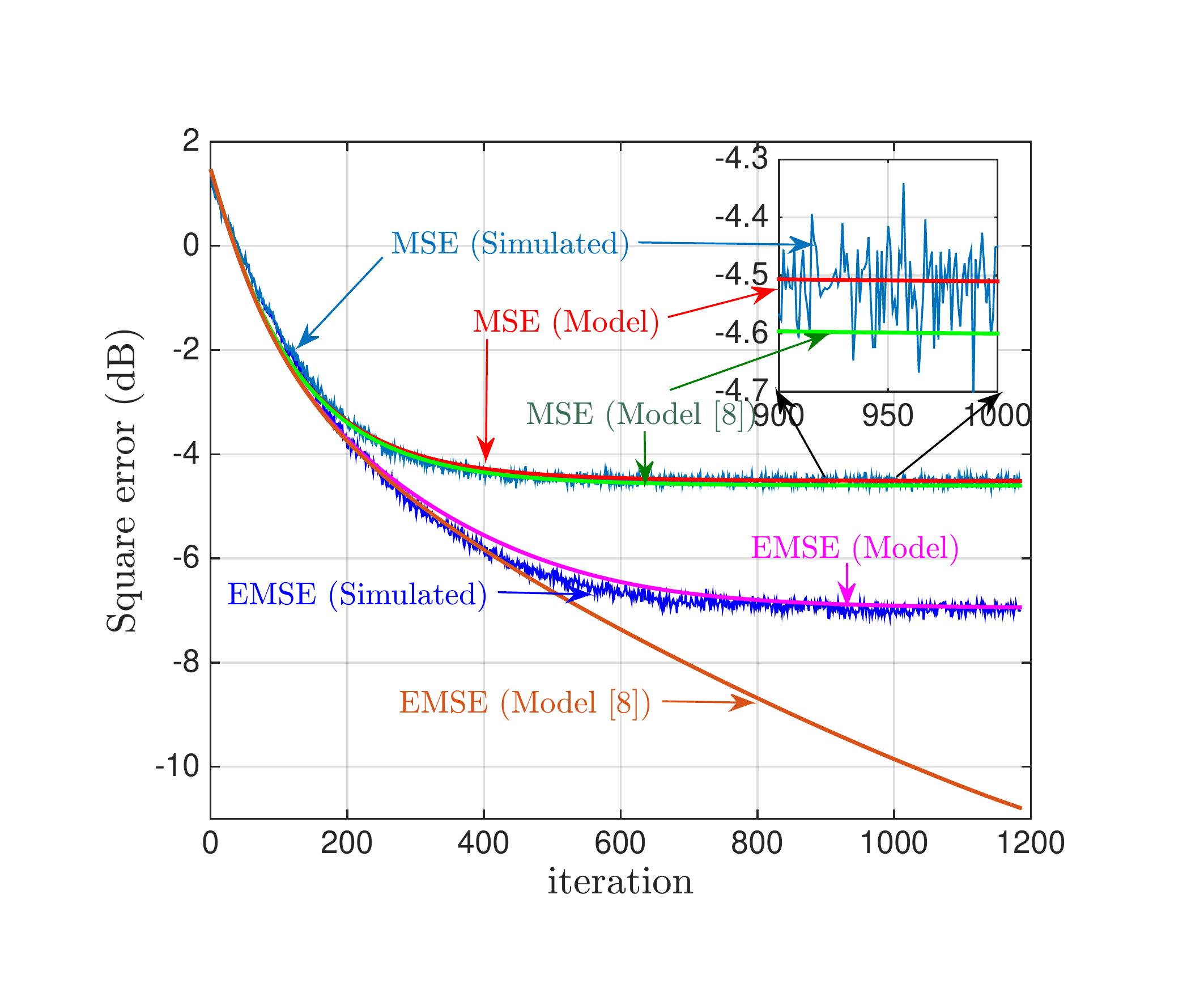} 
          }
	\caption{Model validation and comparison with model obtained in~\cite{Zhang2014}.}
	\label{fig:Model}
	 \vspace{-2mm}
\end{figure*}
 \begin{figure}[!h]
	\subfigure[{$\big[[\tilde{\bw}_n]_3,[\tilde{\bw}_n]_8\big]$} \cwt{dabffcefaef}($n = 800$)]{\label{fig:hist1}	        
		\includegraphics[trim = 38mm 35mm 40mm 35mm, clip, scale=0.18]{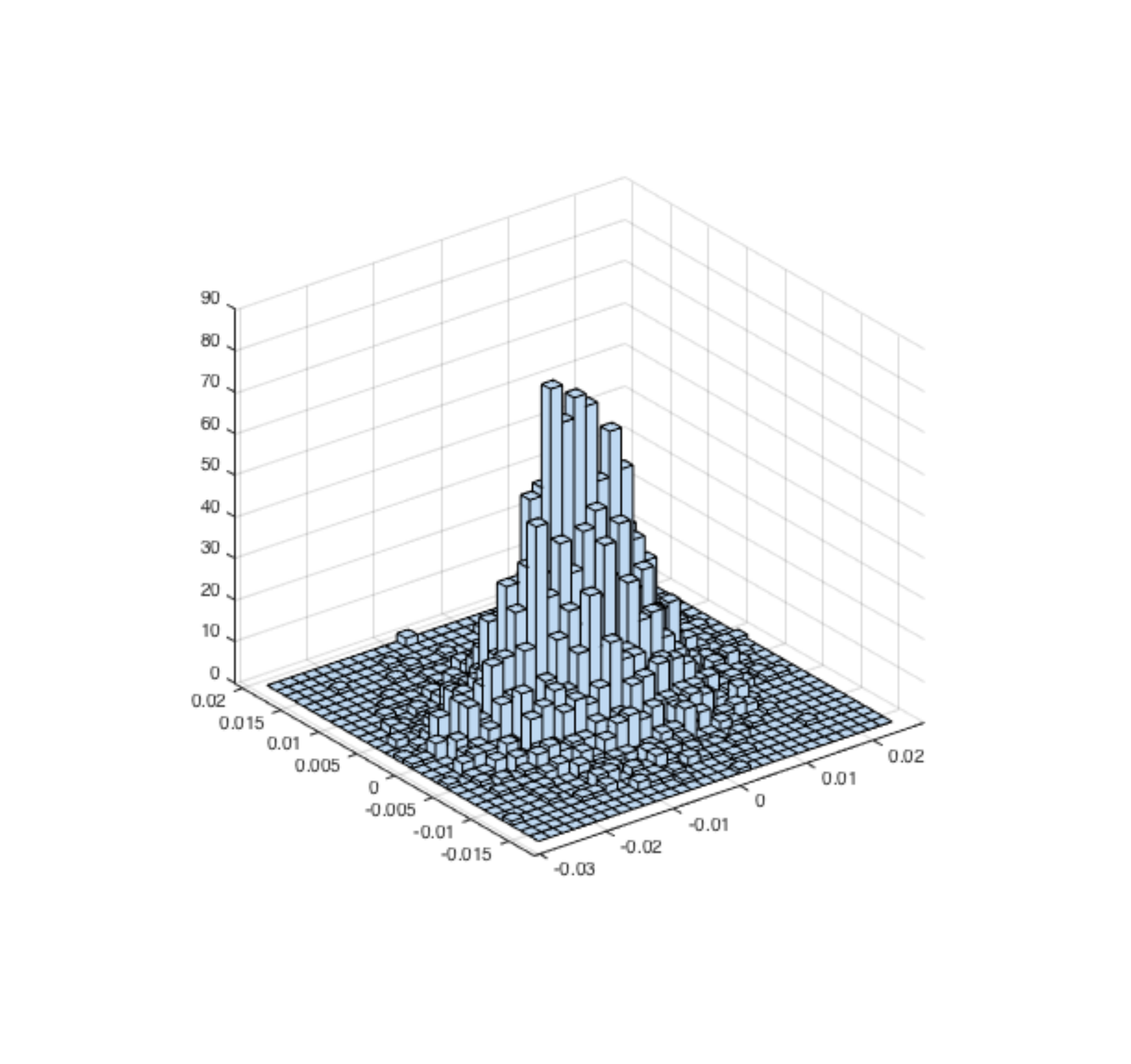} 
          }
          	\subfigure[{$\big[[\tilde{\bw}_n]_3,[\tilde{\bw}_n]_8\big]$}  \cwt{abdesffaef} ($n = 100$)]{ \label{fig:hist2}
		\includegraphics[trim = 38mm 35mm 40mm 35mm, clip, scale=0.18]{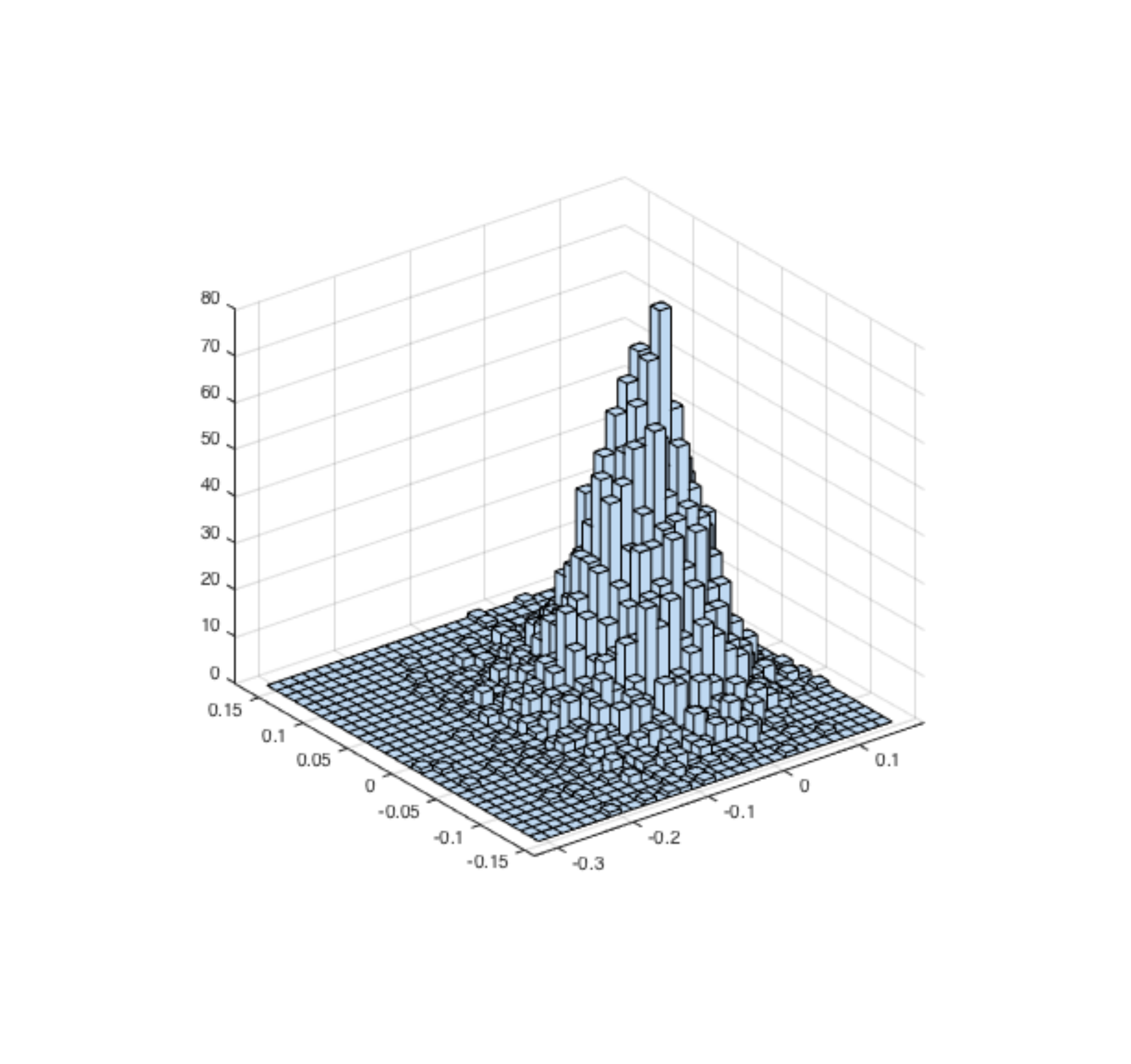} 
          }
          	\subfigure[{$\big[[\tilde{\bw}_n]_9,[\tilde{\bw}_n]_{10}\big]$}  \cwt{abcddefef} ($n = 800$)]{  \label{fig:hist3}
		\includegraphics[trim = 38mm 35mm 40mm 35mm, clip, scale=0.18]{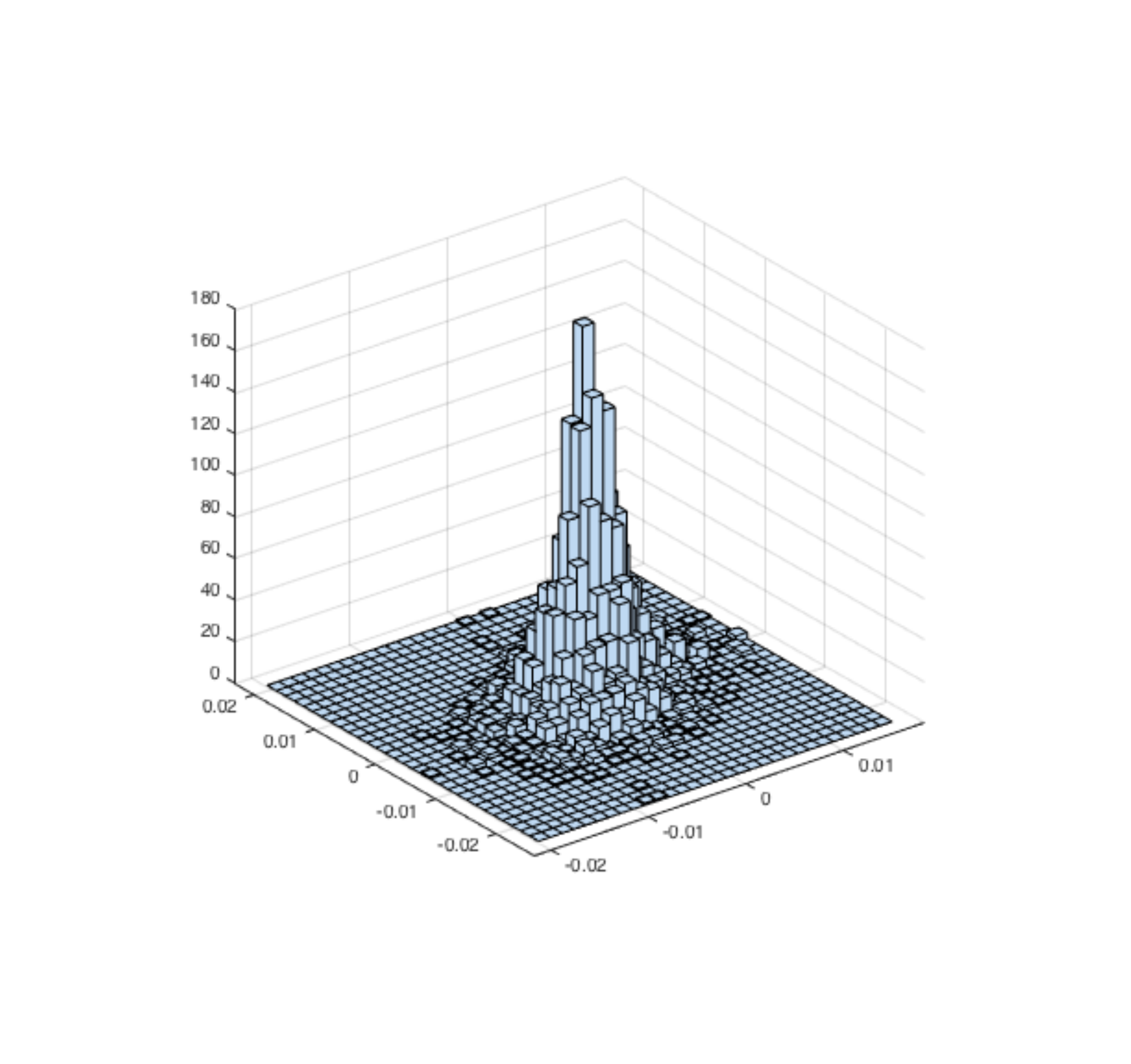} 
          }
	\caption{Histograms of bivariate vector $\big[[\tilde{\bw}_n]_i,[\tilde{\bw}_n]_j\big]$ with 5000 samples.}
	\label{fig:hist}
	 \vspace{-2mm}
\end{figure}
 
\noindent See the proof in Appendix B. The diagonal entries of $\bQ_2$ are simply $[\bQ_2]_{ii} = 1$. The entries $[\bQ_2]_{ij}$ for $i\neq j$ are obtained  by making the following identification:
\begin{align}
	u &\leftarrow {[\bw^\star+\btw_n]_i}  \\
	v &\leftarrow {[\bw^\star+\btw_n]_j}
\end{align}
with 
\begin{equation}
         \mu_u \leftarrow  \E\{ [\bw^\star+\btw_n]_i\} 
 \end{equation}
 \begin{equation}
         \mu_v \leftarrow  \E\{ [\bw^\star+\btw_n]_j\} 
 \end{equation}
\begin{equation}
         \sigma_u^2 \leftarrow  \E\{[\btw_n]_i^2\} - \E\{[\btw_n]_i\}^2  \label{eq:var1} 
 \end{equation}         
\begin{equation}         
         \sigma_v^2  \leftarrow \E\{[\btw_n]_j^2\} - \E\{[\btw_n]_j\}^2  \label{eq:var2 }
 \end{equation}         
 \begin{equation}
         \rho_{uv}  \leftarrow  \E\{[\btw_n]_i[\btw_n]_j\} -  \E\{[\btw_n]_i\}\E\{[\btw_n]_j\}  \label{eq:cov}
 \end{equation}
where $ \E\{[\btw_n]_i[\btw_n]_j\} $ can be extracted from $[\bK_n]_{ij}$.

Considering A1, the term $\bQ_3$ in \eqref{eq:Q3} is given by:
\begin{equation}
	\bQ_3 = \bK_n \bR_x.
\end{equation}

In order to calculate $\bQ_4$, we introduce the following lemma.

\noindent {\lemma Consider two random variables $u$ and $v$ which are jointly Gaussian, namely,
\begin{equation}
	\begin{bmatrix}u \\ v\end{bmatrix} \sim \cp{N} \left( \cb{\mu}:=\begin{bmatrix}\mu_u \\ \mu_v \end{bmatrix} , \, {\bSig}_{uv}:= {\footnotesize
	\begin{bmatrix} \sigma_u^2 & \rho_{uv} \\ \rho_{uv} & \sigma_v^2 \end{bmatrix}} \right) 
\end{equation}
with $\cb{\mu}$ and ${\bSig}_{uv}$ their mean vector and covariance matrix, defined as described above.
The expectation $\E\{u\,\sgn\{v\}\}$ is given by the expressions in Appendix C. \qed}
 
\smallskip
 
It is worth noting that, in Lemma 2 and Lemma 3, the random variables $u$ and $v$ are not necessarily zero-mean. Consequently, the expectations $\E\{\sgn\{u\}\,\sgn\{v\}\}$ and $\E\{u\,\sgn\{v\}\}$ cannot be computed via Price's theorem as in~\cite{sayed2008adaptive,Chen2014NNLMS,Chen.2014.SPL.Steady}. We can compute $\bQ_4$ by making the following identifications:
\begin{align}
        u &\leftarrow  [\btw_n]_i\\
        v &\leftarrow [\bw^\star+\btw_n]_j
\end{align}
with 
\begin{align}
         \mu_u &\leftarrow  \E\{[\btw_n]_i\} \\
         \mu_v &\leftarrow  \E\{ [\bw^\star+\btw_n]_j\} 
\end{align}
The variances and covariance $\sigma_u^2$, $\sigma_v^2$ and $\rho_{uv}$ have the same expressions as in~\eqref{eq:var1}--\eqref{eq:cov} because they are invariant to a constant shift.

The term $\bQ_5$ is then given by:
\begin{equation}
     \bQ_5 = \bR_x\, \bQ_4.
\end{equation}
Replacing $\bQ_1$ to $\bQ_5$ into~\eqref{eq:K.update} we can characterize the second-order behavior of the ZA-LMS algorithm.

\section{Experiment validation}

In this section, we present simulation examples to validate our models. Consider an unknown system of order $L=17$ with weights defined by:
\begin{equation}
\begin{split}
           \bw^\star = [&0.8, \; 0.5, \; 0.3, \; 0.1, \; 0.05, \; \cb{0}_5^\top, \\
            & \,-0.05, \, -0.1, \, -0.3, \, -0.5, \, -0.8]^\top,
\end{split}
\end{equation}
The input signal was a first-order AR process defined as follows: $x_n = 0.6\,x_{n-1} + \nu_{n}$, with $\nu_n$ an i.i.d. zero-mean Gaussian variable with variance $\sigma_\nu^2=0.64$ (so that $\sigma_x^2=1$). The additive noise $z_n$ was zero-mean i.i.d. Gaussian with variance $\sigma_z^2 = 0.01$. The adaptive weights were initialized to $\bw_0=\cb{0}_L$.  The step size was set to $\mu = 0.01$, and the regularization parameter was set to $\lambda = 0.01$ and $0.001$. Note that our objective was not to evaluate the performance of the algorithm but to evaluate the accuracy of our models. The tuning  parameters were thus set arbitrarily. 
 
Before testing our models, we show histograms in Fig.~\ref{fig:hist} to support assumption A2 used in the analysis. They were built from 5000 samples of bivariate vectors $\big[[\tilde{\bw}_n]_3,[\tilde{\bw}_n]_8\big]$ at $n=100$  and $n=800$, and $\big[[\tilde{\bw}_n]_9,[\tilde{\bw}_n]_{10}\big]$ at  $n=800$, for $\mu = 0.01$ and $\lambda = 0.001$. It can be observed that they have Gaussian-like profiles, in particular at steady-state $n=800$. 
 
The accuracy of our models are illustrated in Fig.~\ref{fig:Model}. Simulation results were obtained by averaging over 500 runs. Fig.~\ref{fig:weight} illustrates the mean weight behavior~\eqref{eq:Ebtw.update1} of ZA-LMS with $\lambda = 0.01$. The simulated curves (blue) and theoretical curves~\eqref{eq:Ebtw.update1}--\eqref{eq:lemma1} (red) are superimposed. Fig.~\ref{fig:mse1} shows the MSE and EMSE learning curves~\eqref{eq:MSE}, along with the theoretical curves~\eqref{eq:MSE}--\eqref{eq:Q5} obtained in this paper and in~\cite{Zhang2014} for comparison purpose. For transient MSE, both models are generally consistent with the simulated curve, while a zoom-in on the interval $[900,1000]$ shows that our model has a better consistency. For transient EMSE, our model is consistent with the simulated results while the model proposed in~\cite{Zhang2014} has a large bias. We thus can conclude that in the MSE curve, the inaccuracy of the model in~\cite{Zhang2014} is hidden by the offset $\sigma_z^2$ in~\eqref{eq:MSE} that is much higher than EMSE. Fig.~\ref{fig:mse2} shows a similar result with $\lambda = 0.001$.

\section{Conclusion}
In this paper, we derived analytical models to characterize the transient behavior of ZA-LMS in the mean and mean-square sense. Simulations illustrated the consistency between Monte Carlo learning curves and the theoretical findings, as well as the improved accuracy of our model compared with previous works. Extension to more general update relation such as the reweighted ZA-LMS algorithm will be considered in future works.

\section*{Appendix A: proof of Lemma 1} 
\noindent Consider a Gaussian random variable $u\sim\cp{N}(\mu, \sigma^2)$. We have:
\begin{equation}
       \begin{split}
           &\E\{\sgn\{u\}\} \\ 
		&= \int_{-\infty}^{0} - \cp{N}(\mu, \sigma^2) du +  \int_{0}^{+\infty}  \cp{N}(\mu, \sigma^2) du \\
		&= 1-2\phi(-{\mu}/{\sigma}).
       \end{split}
\end{equation}
\qed

\section*{Appendix B: proof of Lemma 2} 
\noindent Consider jointly Gaussian variables $u$ and $v$. We have:
\begin{equation}
      \begin{split}
            &\E\{\sgn\{u\}\sgn\{v\}\} \\
              =&  \int_{-\infty}^{0}  \int_{-\infty}^{0} \cp{N}([u,v]^\top, \bSig_{uv})\, du \,dv   \\
               & + \int_{0}^{+\infty}  \int_{0}^{+\infty} \cp{N}([u,v]^\top, \bSig_{uv})\, du \,dv  \\
               & -  \int_{-\infty}^{0} \int_{0}^{+\infty} \cp{N}([u,v]^\top, \bSig_{uv})\, du\, dv  \\
               & - \int_{0}^{+\infty}\int_{-\infty}^{0}\cp{N}([u,v]^\top, \bSig_{uv})\, du \, dv.                  
            \end{split}
\end{equation}
With simple algebraic manipulations and using the definition of CDF of the multivariate Gaussian distribution, we write: 
\begin{equation}
	\begin{split}
	&\E\{\sgn\{u\}\sgn\{v\}\} \\
	& = \Phi(\cb{0}_2, [\mu_u, \mu_v]^\top, \bSig_{uv})+ \Phi(\cb{0}_2, -[\mu_u, \mu_v]^\top, \bSig_{uv}) \\
	& -  \Phi(\cb{0}_2, [\mu_u, -\mu_v]^\top, \overline{\bSig}_{uv}) -  \Phi(\cb{0}_2, [-\mu_u, \mu_v]^\top, \overline{\bSig}_{uv})
	\end{split}
\end{equation}
where $\overline{\bSig}_{uv} = {\bSig}_{uv} \circ {\footnotesize \begin{bmatrix}  \phantom{-}1  & -1\\ -1 & \phantom{-}1 \end{bmatrix}} $.
\qed

\section*{Appendix C: proof of Lemma 3} 
\noindent Consider jointly Gaussian variables $u$ and $v$. We have:
\begin{equation}
       \label{eq:usignv}
      \begin{split}
            &\E\{u\,\sgn\{v\}\} \\
             &=  \int_{-\infty}^{+\infty}   \sgn\{v\} \left( \int_{-\infty}^{+\infty} {u}\, \cp{N}([u,v]^\top, \bSig_{uv})\, du \right) dv.  \\              
            \end{split}
\end{equation}
For ease of presentation, we write:
\begin{equation}
        \bSig_{uv}^{-1} =  { \begin{bmatrix}  a  & c\\ c & b \end{bmatrix}}.
\end{equation}
Since $\bSig_{uv}$ is positive definite, we have $a>0$ and $ab-c^2 > 0$. This remark is important since $b-\frac{c^2}{a}$ will denote below the variance of a Gaussian distribution.
The integral~\eqref{eq:usignv} can be expressed via the above quantities:
\begin{equation}
     \label{eq:Eint}
     \begin{split}
     \eqref{eq:usignv} &= \frac{1}{2\pi \sqrt{|\bSig_{uv}|}}  \int_{-\infty}^{+\infty} \left\{ \sgn\{v\} \exp\!\Big[\!-\!\frac{1}{2}(b-\frac{c^2}{a})(v-\mu_v)^2\Big]
     \right. \\
                                 &\times \bigg(\sqrt{\frac{2\pi}{a}}\int_{-\infty}^{+\infty} u \, {\textstyle{\cp{N}(\mu_u-\frac{c}{a}(v-\mu_v),\frac{1}{a})}}\, du\bigg)
                                 \bigg\} \, dv         \\     
                                 &= \frac{1}{\sqrt{2\pi a|\bSig_{uv}|}} \int_{-\infty}^{+\infty}\Big\{ \sgn\{v\} \exp\!\Big[\!-\!\frac{1}{2}(b-\frac{c^2}{a})(v-\mu_v)^2\Big] \\
                                 &\times \left(\mu_u-\frac{c}{a}(v-\mu_v)\right)\Big\}\,dv \\
                                 &=\frac{1}{\sqrt{2\pi a|\bSig_{uv}|}} \\
                                 &\times\Big\{\int_{-\infty}^{+\infty} \!\!\left(\mu_u+\frac{c}{a}\mu_v \right) \sgn\{v\} \exp\!\Big[\!-\!\frac{1}{2}\delta(v-\mu_v)^2\Big]\, dv \\
                                 &-   \int_{-\infty}^{+\infty} \!\! \frac{c}{a}\, v\,\sgn\{v\} \exp\!\Big[\!-\!\frac{1}{2}\delta(v-\mu_v)^2\Big] \,dv\Big\}
     \end{split}
\end{equation}
where $\delta = b-\frac{c^2}{a}>0$. The second equality comes from the fact that the second integral in the first equality is the mean value of $\cp{N}(\mu_u-\frac{c}{a}(v-\mu_v),\frac{1}{a})$. Using Lemma~1, we get:  
\begin{equation}
	\label{eq:int1}
	\begin{split}
	&\int_{-\infty}^{+\infty} \!\!\left(\mu_u+\frac{c}{a}\mu_v \right) \sgn\{v\} \exp\!\Big[\!-\!\frac{1}{2}\delta(v-\mu_v)^2\Big]\, dv \\
	&= \left(\mu_u+\frac{c}{a}\mu_v \right){\sqrt{\frac{2\pi}{\delta}}}\Big[1 - 2 \phi\Big(\mu_v \sqrt{\delta}\Big)\Big].
	\end{split}
\end{equation}
Finally, by considering the mean value of the folded Gaussian distribution~\cite{Leone1961}, we have:
\begin{equation}
	\label{eq:int2}
	\begin{split}
	&\int_{-\infty}^{+\infty} \!\! \frac{c}{a}\, v\,\sgn\{v\} \exp\!\Big[\!-\!\frac{1}{2}\delta(v-\mu_v)^2\Big] \,dv \\
	&= \frac{c}{a}\sqrt{\frac{2\pi}{\delta}}   \int_{-\infty}^{+\infty}  |v|\, \frac{1}{\sqrt{2\pi}\sqrt{{1}/{\delta}}}\, \exp\!\Big[\!-\!\frac{(v-\mu_v)^2}{2 \, ({1}/{\delta})}\Big] dv \\
         &= \frac{c}{a}\sqrt{\frac{2\pi}{\delta}} \left[\sqrt{\frac{2}{\delta\pi}} \exp\!\Big(\!-\frac{1}{2}\mu_v^2\,\delta\Big) 
         + \mu_v\Big(1-2\phi\big(-\mu_v \sqrt{\delta}\,\big)\Big)\right].
       \end{split}
\end{equation}
Substituting the results in~\eqref{eq:int1} and~\eqref{eq:int2} into~\eqref{eq:Eint} yields the value of $\E\{u\,\sgn\{v\}\}$.
\qed

\bibliographystyle{IEEEbib}
\bibliography{ref}

\begin{thebibliography}{10}

\bibitem{bruckstein2009sparse}
A.~M. Bruckstein, D.~L. Donoho, and M.~Elad,
\newblock ``From sparse solutions of systems of equations to sparse modeling of
  signals and images,''
\newblock {\em SIAM review}, vol. 51, no. 1, pp. 34--81, 2009.

\bibitem{Elad2010sparse}
M.~Elad,
\newblock {\em Sparse and Redundant Representations: From Theory to
  Applications in Signal and Image Processing},
\newblock Springer, 2010.

\bibitem{fessler2010model}
J.~A. Fessler,
\newblock ``Model-based image reconstruction for {MRI},''
\newblock {\em IEEE Signal Processing Magazine}, vol. 27, no. 4, pp. 81--89,
  July 2010.

\bibitem{iordache2011sparse}
M.~D. Iordache, J.~M. Bioucas-Dias, and A.~Plaza,
\newblock ``Sparse unmixing of hyperspectral data,''
\newblock {\em IEEE Transactions on Geoscience and Remote Sensing}, vol. 49,
  no. 6, pp. 2014--2039, June 2011.

\bibitem{cotter2002sparse}
S.~F. Cotter and B.~D. Rao,
\newblock ``Sparse channel estimation via matching pursuit with application to
  equalization,''
\newblock {\em IEEE Transactions on Communications}, vol. 50, no. 3, pp.
  374--377, 2002.

\bibitem{candes2008introduction}
E.~J. Cand\`es and M.~B. Wakin,
\newblock ``An introduction to compressive sampling,''
\newblock {\em IEEE Signal Processing Magazine}, vol. 25, no. 2, pp. 21--30,
  2008.

\bibitem{Chen2009ZALMS}
Y.~Chen, Y.~Gu, and A.~O. Hero,
\newblock ``Sparse {LMS} for system identification,''
\newblock in {\em Proc. IEEE Int. Conf. Acoust., Speech, Signal Process.
  (ICASSP)}, Taipei, China., Apr. 2009, pp. 3125--3128.

\bibitem{Zhang2014}
S.~Zhang and J.~Zhang,
\newblock ``Transient analysis of zero attracting {NLMS} algorithm without
  {Gaussian} inputs assumption,''
\newblock {\em Signal processing}, vol. 97, pp. 100--109, Apr. 2014.

\bibitem{Shi2010}
K.~Shi and P.~Shi,
\newblock ``Convergence analysis of sparse {LMS} algorithms with $l_1$-norm
  penalty based on white input signal,''
\newblock {\em Signal processing}, vol. 90, no. 12, Dec. 2010.

\bibitem{Su2012}
G.~Su, J.~Jin, Y.~Gu, and J.~Wang,
\newblock ``Performance analysis of $\ell_0$ norm constraint least mean square
  algorithm,''
\newblock {\em IEEE Transactions on Signal Processing}, vol. 60, no. 6, pp.
  2223--2235, May 2012.

\bibitem{Chen2014NNLMS}
J.~Chen, C.~Richard, J.-C.~M. Bermudez, and P.~Honeine,
\newblock ``Variants of non-negative least-mean-square algorithm and
  convergence analysis,''
\newblock {\em IEEE Transactions on Signal Processing}, vol. 62, no. 15, pp.
  3990--4005, Aug. 2014.

\bibitem{haykin2005}
S.~Haykin,
\newblock {\em Adaptive Filter Theory},
\newblock Pearson Education India, 4th edition, 2005.

\bibitem{sayed2008adaptive}
A.~H. Sayed,
\newblock {\em Adaptive {F}ilters},
\newblock John Wiley \& Sons, 2008.

\bibitem{Chen.2014.SPL.Steady}
J.~Chen, C.~Richard, and J.-C.~M. Bermudez,
\newblock ``Steady-state performance of non-negative least-mean-square
  algorithm and its variants,''
\newblock {\em IEEE Signal Processing letters}, vol. 21, no. 8, pp. 928--932,
  Aug. 2014.

\bibitem{Leone1961}
F.~C. Leone, R.~B. Nottingham, and L.~S. Nelson,
\newblock ``The folded normal distribution,''
\newblock {\em Technometrics}, vol. 3, no. 4, pp. 543--550, 1961.

\end{thebibliography}
\balance

\end{document}